\documentclass[fleqn,usenatbib]{mnras}

\usepackage{amsfonts,amsmath,amssymb,bm,booktabs,cleveref,dcolumn,enumitem,graphicx,multirow,nomencl,physics,svg,xcolor,arydshln,float,balance}

\usepackage[version=4]{mhchem} 

\usepackage[skip=0pt]{caption}

\usepackage{orcidlink}

\graphicspath{{Figures/}}

\makenomenclature

\newcolumntype{d}{D{.}{.}{-1}}
\newcolumntype{H}{>{\setbox0=\hbox\bgroup}c<{\egroup}@{}}

\newcommand{\Abinitio}{\emph{Ab initio}}

\newcommand{\abinitio}{\emph{ab initio}}

\newcommand{\eg}{\emph{e.g.}}

\newcommand{\cm}{cm$^{-1}$}
\newcommand{\um}{$\mu$m}

\newcommand{\Duo}{Duo}
\newcommand{\DUO}{Duo}
\newcommand{\STATES}{.states} 
\newcommand{\TRANS}{.trans}
\newcommand{\MODEL}{.model}
\newcommand{\EXOCROSS}{ExoCross}
\newcommand{\EXOMOL}{ExoMol}

\newcommand{\Marvel}{MARVEL}
\newcommand{\MARVEL}{MARVEL}
\newcommand{\Mollist}{MoLLIST}
\newcommand{\MOLLIST}{MoLLIST}

\newcommand{\MOLPRO}{Molpro}

\newcommand{\LLname}{{\sc kNigHt}}
\newcommand{\NHbold}{$^\mathbf{14}$N$^\mathbf{1}$H}
\newcommand{\NHl}{NH}

\newcommand{\NH}{$^{14}$N$^{1}$H}
\newcommand{\isoa}{$^{14}$N$^{2}$H}
\newcommand{\isob}{$^{15}$N$^{1}$H}
\newcommand{\isoc}{$^{15}$N$^{2}$H}

\newcommand{\NHX}{X~$^3\Sigma^-$}
\newcommand{\NHA}{A~$^3\Pi$}
\newcommand{\NHaa}{a~$^1\Delta$}
\newcommand{\NHbb}{b~$^1\Sigma^+$}
\newcommand{\NHcc}{c~$^1\Pi$}

\newcommand{\noenergies}{4,076}
\newcommand{\noMaenergies}{1,078}
\newcommand{\noMaenergiesunique}{1,023}
\newcommand{\noEHenergies}{76}
\newcommand{\noCaenergies}{2,922}
\newcommand{\notransitions}{327,014}

\newcommand{\noMafitting}{1,053}

\defcitealias{Huber1979MolecularMolecules}{H \& H (1979)}
\defcitealias{19DarbyNH}{\MARVEL{}}
\defcitealias{Hammer1979}{H \& D (1979)}
\defcitealias{Barklem2016}{B \& C (2016)}

\title[\LLname{} line list for NH]{Full spectroscopic model and trihybrid experimental-perturbative-variational line list for NH}

\author[A. N. Perri and L. K. McKemmish]
{Armando N. Perri\orcidlink{0009-0005-1737-2569} and Laura K. McKemmish\orcidlink{0000-0003-1039-2143}\thanks{E-mail: l.mckemmish@unsw.edu.au}
\vspace*{6pt}\\
School of Chemistry, University of New South Wales, 2052, Sydney, Australia\\}

\date{Accepted XXX. Received YYY; in original form ZZZ}

\pubyear{2024} 

\begin{document}

\date{\today}

\maketitle

\begin{abstract}

Imidogen (NH) is a reactive molecule whose presence in astrochemical environments is of interest due to its role in the formation of nitrogen-containing molecules and as a potential probe of nitrogen abundance. Spectroscopic NH monitoring is useful for Earth-based combustion and photolysis processes of ammonia and other nitrogen-containing species. NH is also relevant to ultracold molecular physics and plasma studies. To enable these diverse applications, high-quality molecular spectroscopic data is required. Here, a new line list with significant advantages over existing data is presented. Most notably, this line list models isotopologue spectroscopy and forbidden transitions (important for NH visible absorption), alongside some overall improvements to accuracy and completeness. This approach takes advantage of existing experimental data (from a previous \Marvel{} compilation) and perturbative line lists together with new MRCI \abinitio{} electronic data. These are used to produce a novel variational spectroscopic model and trihybrid line list for the main \NH{} isotopologue, as well as isotopologue-extrapolated hybrid line lists for the \isoa{}, \isob{} and \isoc{} isotopologues. The new \NH{} \EXOMOL{}-style trihybrid line list, \LLname{}, comprises \noenergies{} energy levels (\noMaenergies{} experimental) and \notransitions{} transitions up to 47,500~\cm{} (211~nm) between five low-lying electronic states (\NHX{}, \NHaa{}, \NHbb{}, \NHA{} and \NHcc{}). For most anticipated applications aside from far-IR studies, this line list will be of sufficient quality; any improvements should focus on the \NHbb{} energies, and the \NHaa{}~--~\NHA{} and \NHbb{}~--~\NHA{} spin-orbit couplings. 
\end{abstract}


\section{Introduction}

Imidogen (NH) is a reactive free radical diatomic molecule that is likely the simplest nitrogen-containing compound in the universe. Its spectroscopic properties, however, make it quite challenging to observe in astronomical environments. Nevertheless, NH has been observed in a wide variety of systems, as recently discussed in \cite{civivs2023infrared}. Briefly, as a non-exhaustive list of observational studies, NH has been detected in the atmosphere of the Sun \citep{Babcock1933CHEMICALSUN,grevesse1990identification,civivs2023infrared}, the interstellar medium \citep{Meyer1991DiscoveryNH,crawford1997detection,cernicharo2000far,weselak2009interstellar,persson2010nitrogen,persson2012nitrogen}, comets \citep{Swings1941THE1940c} and cool stars \citep{sneden1973nitrogen, lambert1984carbon, smith1986chemical, Aoki19971997AA___328__175A}. Many of these studies use NH as a proxy to quantify nitrogen abundance in a way that is independent of other elemental abundances (assuming that hydrogen is in excess).  

As a reactive free radical, NH plays an important chemical role in the astrochemical production of molecules, with current chemical models incorporating NH as a reactant or intermediate, for example, in the formation of \ce{N2} \citep{C5CP05190H}, ammonia \citep{Dustgrains1} and even the DNA nucleotide base cytosine \citep{GUPTA2013797}. Based on the observed abundance of NH in interstellar space, the formation of NH is generally accepted to occur on dust grains \citep{Dustgrains1,Wagenblast1993OnClouds}, although radiative association of N and H atoms could contribute in low-dust environments \citep{szabo2019formation}. 

NH is also a prominent reactive intermediate in many non-astronomical contexts and thus spectroscopic monitoring of NH can provide insights into reaction pathways in varied environments. For example, NH is monitored to understand the chemistry of ammonia photolysis (an important sink in Earth's atmosphere) \citep{flash2} and various plasma systems \citep{pflieger2017use,Hamdan2018MicrowaveTreatment,Harilal2018AnPlasma}. The chemistry of ammonia and other nitrogen-containing compounds during combustion, understood through NH spectroscopy, thus holds significant contemporary relevance \citep{GlarborgModelingCombustion,brackmann2018formation,lamoureux2019quantitative,jin2022experimental,mashruk2023combustion}. Further, ammonia is emerging as a promising alternative fuel source. It is imperative, however, to prevent incomplete combustion in order to avoid the undesirable generation of NOx pollutants.  

At the other end of the temperature scale, NH is of interest for experimental investigations within ultracold molecular physics, with various efforts proposed or demonstrated for cooling. Specifically, NH has already been cooled with buffer gases into a magnetic trap \citep{Campbell2007MagneticX-3}, decelerated using the Stark effect \citep{van2006production} and Zeeman effect \citep{plomp2019multistage}. NH has also been proposed as a candidate for laser cooling \citep{Li2021ExcellentInterfere,yan2021direct} due to the similarities in the \NHA{} and \NHX{} potential energy surfaces leading to a near-diagonal \NHA{}~--~\NHX{} transition with a Frank-Condon factor of 0.999472 \citep{Liu2020TheDatabase}, which minimises loss mechanisms in the laser cooling process.

As alluded to above, despite its simplicity, the position and strength of the spectral bands of NH can make it a challenging molecule to detect astrophysically. The visible region only has weak spectral features from forbidden transitions or vibrationally excited initial states with low Boltzmann population. In the literature, NH is most commonly detected astronomically through its strong ultraviolet bands at approximately 335~nm, but ground-based observations suffer from significant atmospheric molecular absorption in this window \citep{weselak2009interstellar}. Since NH is such a light molecule, the pure rotational bands move into the far infrared spectral region ($\approx$~1~THz) accessible by few telescopes; the first observation of this band (according to the thorough literature review of \cite{mcguire20222021}) was in 2000 by \cite{cernicharo2000far} using Infrared Space Observatory (in operation 1995~--~1998). The mid-infrared vibrational transitions are allowed and span a considerable spectral range for hot stars, but ground-based observations do need to be cautious of telluric absorption.  New and emerging opportunities with the next generation in space-based infrared telescopes starting with JWST mean that the detection of NH in this region is becoming easier. 

In order to understand and observe molecules in space, astronomers rely on line lists, that is, the set of molecular energies and transition intensities between these energy levels; this data is a crucial input to models of astronomical objects and necessary to interpret observational spectroscopically-resolved data. For NH, Bernath and co-workers have constructed the \MOLLIST{} line list using the traditional approach (based on an effective Hamiltonian treatment), with \NHX{} rovibrational and rotational transitions considered in \cite{Brooke2014LineNH,Brooke2015Note:NH} and \NHA{}~--~\NHX{} rovibronic transitions modelled in \cite{18FernandoNH}. These data were made available in the conventional \EXOMOL{} format in \cite{Villanueva2018PlanetaryExoplanets}.

Despite this, there are a number of opportunities to improve these spectral data, which have been acted upon herein to produce \LLname{}, the best line list currently available for \NH{} and its isotopologues across the UV to mid-IR regions (studies of the far-IR region should instead utilise the Splatalogue database \citep{Splatalogue}, which includes hyperfine-resolved rotational data from JPL \citep{JPL} and CDMS \citep{CDMS}). Importantly, the trihybrid approach used in this paper combines experimental (\MARVEL{}) data, perturbative (\MOLLIST{}) data and novel variational data from a fitted spectroscopic model (potential energy, dipole moment and coupling curves) to produce a line list that has both high accuracy of strong line positions and high completeness (for opacity calculations and radiative transfer modelling). This approach has been previously used in the line list construction of CN \citep{Syme2021FullCN} and ZrO \citep{Perri2023} in accordance with the \EXOMOL{} project \citep{Tennyson2012ExoMol:Atmospheres, Tennyson2020}. It should be noted that accurate variational calculations for diatomic molecules using global coupled channels (\eg{} \cite{1999Amiot, LETELIER20061030, B700044H, LEROY2017167}) have proved successful for many decades.

\begin{figure}
\centering
\includegraphics[width=\linewidth]{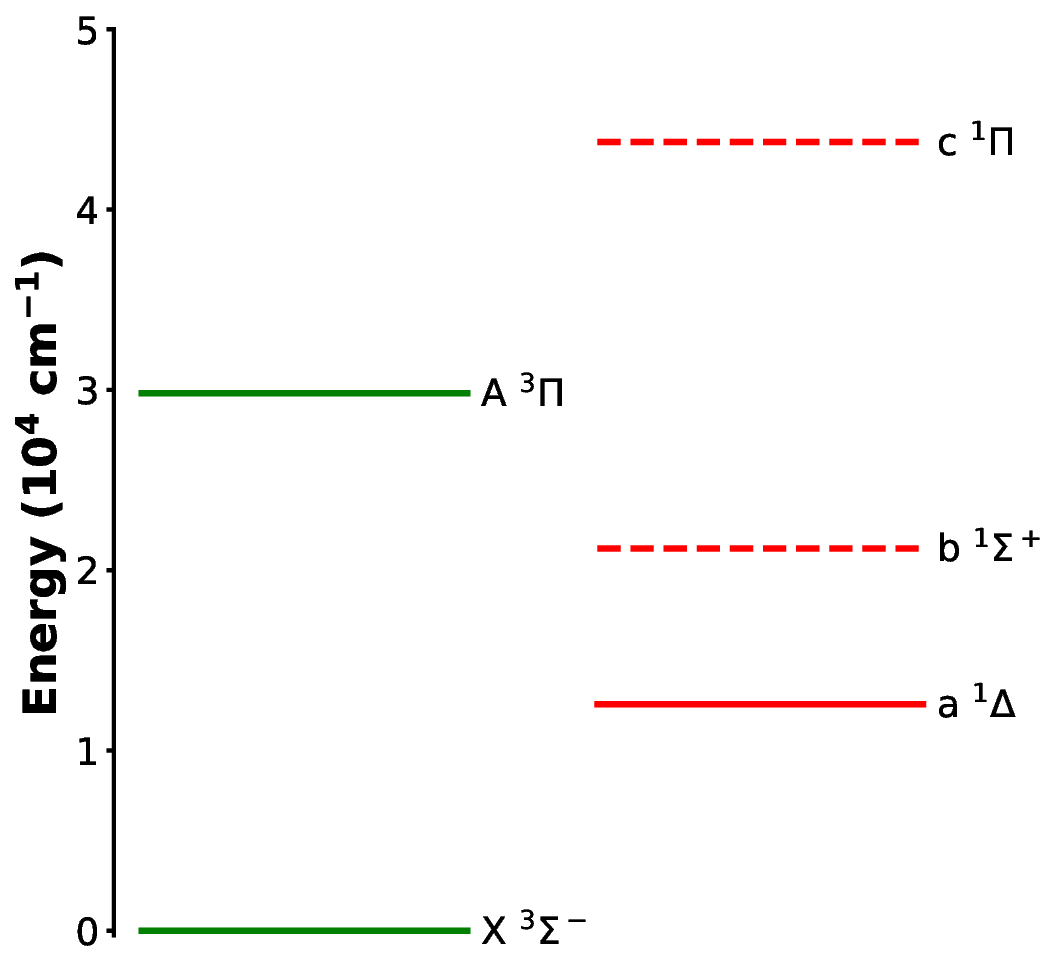}
\caption{The five lowest electronic states of NH with term energies $T_e$ from Huber \& Herzberg (1979). The solid lines indicate electronic states fit to experimental data in this work.}
\label{f:NH_states}
\end{figure}

Specifically, the new \LLname{} line list improves the existing \Mollist{} data in three ways. Firstly, it includes additional singlet electronic states and spin-orbit coupling between states, which allow for the spin-forbidden \NHbb{}~--~\NHX{} and \NHaa{}~--~\NHX{} visible transitions \citep{yarkony1989electronic} to be modelled. These forbidden transitions are actually the dominant absorption source over almost all of the visible spectral region. Second, it incorporates the recently compiled comprehensive and self-consistent \Marvel{} energy levels for \NH{}, which include almost all experimental assigned transitions (see \cite{19DarbyNH} for details) in order to increase the quality of the predicted transition frequencies. The nature of this combined experimental analysis has allowed for unprecedented accuracy of spectral line positions with reliable quantum number assignment \citep{Flaud1976,07FurtenbacherMARVEL}. Finally, the variational approach based on a fitted spectroscopic model employed here trivially allows the generation of isotopologue spectra that are challenging to obtain using traditional perturbative approaches. 

The new \LLname{} line list could also be of significant interest to the ultracold molecular physics community, where these data could be used to model the laser cooling mechanism in fine detail to quantify all loss mechanisms including intersystem crossing. The effects of the \NHaa{}~--~\NHA{} and \NHbb{}~--~\NHA{} spin-orbit couplings quantified in this work may be particularly useful, especially given the \abinitio{} spin-orbit coupling curves that are calculated. The limitations of the \LLname{} line list are explicitly discussed and future potential areas for improvement are identified.

This article is structured as follows; an overview of the literature experimental, perturbative and \abinitio{} spectroscopic data are presented in Section~\ref{s:abinitio_NH}. \Abinitio{} data are presented in this section. In Section~\ref{s:NHSM}, the spectroscopic model for the main \NH{} isotopologue is constructed based on \abinitio{} curves fitted to experimental data. In Section~\ref{s:llc}, variational line lists are generated for the \NH{}, \isoa{}, \isob{} and \isoc{} isotopologues, as well as trihybrid and isotopologue-extrapolated hybrid line lists. In Section~\ref{s:lla}, partition functions and cross sections are analysed. The accuracy of the total \NH{} cross section is considered in the context of astronomical observations from both JWST and similar space-based telescopes, as well as for high-resolution cross-correlation methodologies using ground-based telescopes.

\section[\NH{} Spectroscopic Data]{\NHbold{} Spectroscopic Data}\label{s:abinitio_NH}

The rovibronic spectroscopy of NH is characterised primarily by the five electronic states in \Cref{f:NH_states} \citep{OwonoOwono2007TheoreticalFunctions,Song2016AccurateEffect}. The ground $1\sigma^22\sigma^23\sigma^21\pi^2$ electronic configuration gives the ground \NHX{} electronic state, as well as the low-lying excited \NHaa{} and \NHbb{} electronic states. Excitation into the $1\sigma^22\sigma^23\sigma^11\pi^3$ electronic configuration produces the \NHA{} and \NHcc{} electronic states. The \NHA{} -- \NHX{} rovibronic system dominates the spectrum, however, the spin-forbidden \NHaa{} -- \NHX{} and \NHbb{} -- \NHX{} may be observed over the \NHX{} rovibrational system in the visible region due to \NHaa{} -- \NHA{} and \NHbb{} -- \NHA{} spin-orbit couplings. As the Boltzmann population of the \NHaa{} electronic state is negligible even at high temperatures, transitions between singlet states are not important to the spectroscopy of thermally-equilibrated NH.

\subsection{Experimental data}
In the literature, the main \NH{} isotopologue has been~investigated through many experimental spectroscopic studies. This was recently reviewed and collated in \cite{19DarbyNH}, which obtained assigned rovibronic transitions from eighteen publications \citep{82BeAm, 82RaSa, 82VaMeDy, 85HaAdKaCu, 86BoBrChGu, Brazier1986FourierNH, 86LeEvBr, 86RaBe, 86UbMeTeDy, 90HaMi, 91GeSaGrFa, 97KlTaWi, 99RaBeHi, 03VaSaMoJo, 04FlBrMaOd, 04LeBrWiSi, 07RoBrFlZi, Ram2010RevisedNH} derived from both laboratory measurements and astronomical spectra. Using the \MARVEL{} \citep{07FurtenbacherMARVEL} algorithm, the 3,002 rovibronic transitions validated were inverted to give a set of empirical energies for rovibronic states belonging to the \NHX{}, \NHA{}, \NHaa{} and \NHcc{} electronic states. These energies form one self-consistent spectroscopic network containing 1,058 rovibronic states. Apart from unresolved $\Lambda$-doubling in the \NHaa{} electronic state, all fine structure is characterised using Hund's coupling case (b). The \NHbb{} electronic state data were found to be sparse and thus not included in the study.

\begin{figure}
\centering
\includegraphics[width=\linewidth]{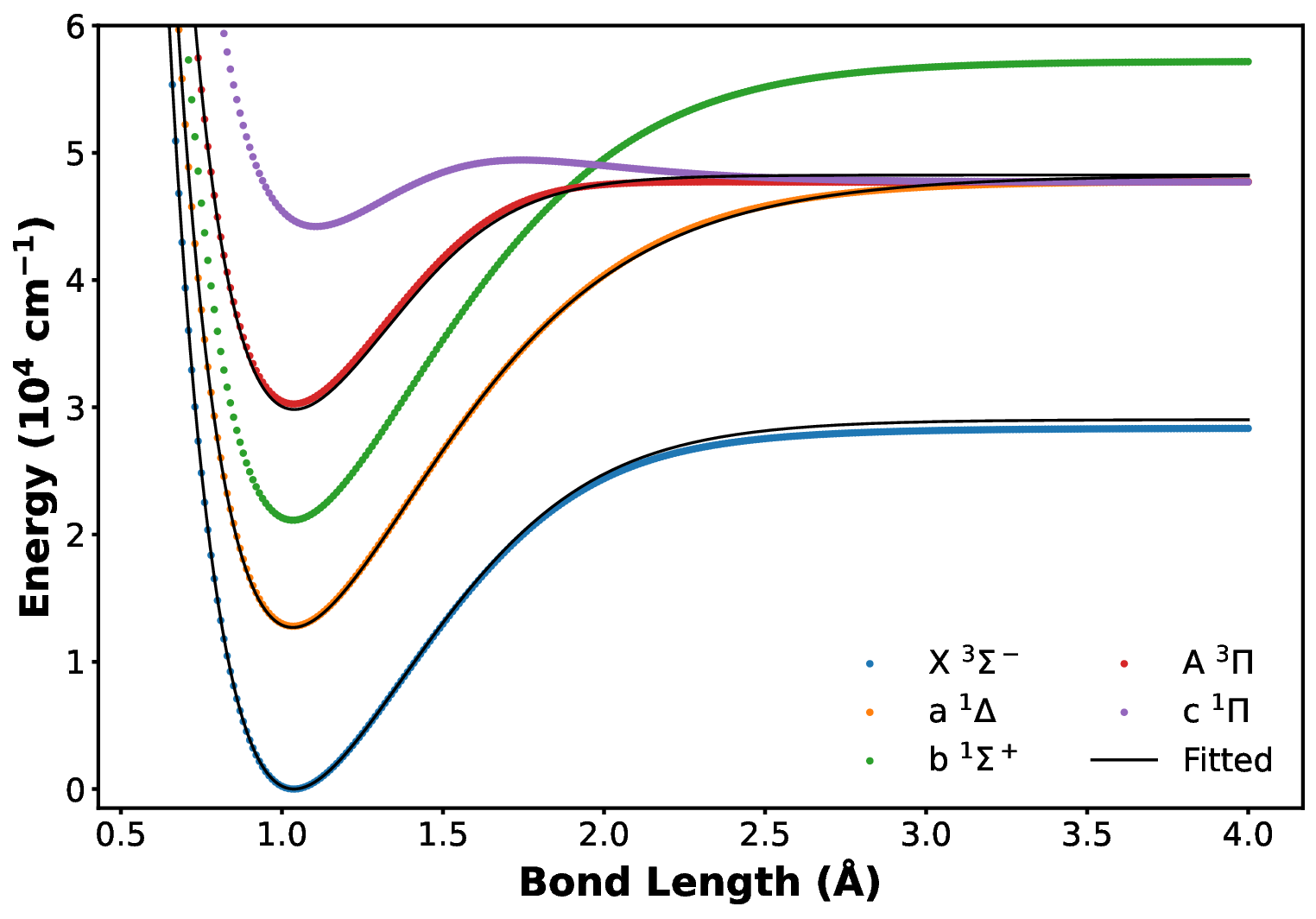}
\caption{The potential energy curves of the five lowest electronic states of \NH{}. The coloured dots show the \abinitio{} data calculated in this work. The solid black lines illustrate the final curves fit for the \NHX{}, \NHaa{} and \NHA{} electronic states, which were modelled using an extended Morse oscillator curve fit to the \MARVEL{} experimental data.}
\label{f:PECNH}
\end{figure}

\subsection{Perturbative line lists}
Beyond the experimental data collated by \cite{19DarbyNH}, there exists only perturbative line lists for NH~in~the literature. The perturbative (or traditional) methodology relies on an effective Hamiltonian, where experimental data are fit to polynomial expressions through the optimisation of spectroscopic parameters for each electronic state. This approach allows for the prediction of unobserved transitions with \abinitio{} dipole moment curves. This provides a natural extension to experimental data, where the associated line lists interpolate experimental energies to high accuracy. Due to corrections obtained through perturbation theory, however, perturbative line lists typically extrapolate poorly, especially to higher vibrational levels. For this reason, they~are often incomplete as they are limited by experimental data.

Perturbative line lists for \NH{} were generated by both \cite{Brooke2014LineNH,Brooke2015Note:NH} and \cite{18FernandoNH}. The former line list contains \NHX{} rovibrational and rotational transitions with fine structure characterised using Hund's coupling case (b). The Einstein A coefficients were derived from an \abinitio{} diagonal dipole moment curve calculated using the MRCI/aug-cc-pV6Z model chemistry. These transition intensities were revised in \cite{Brooke2015Note:NH} using an improved method of converting the transition dipole matrix elements from Hund's coupling case (b) to (a) as required by PGOPHER \citep{Western2017PGOPHER:Spectra}. The latter line list specifies \NHA{} -- \NHX{} rovibronic transitions with fine structure characterised using Hund's coupling case (b). The off-diagonal dipole moment curve was calculated using the MRCI/aug-cc-pwCV5Z model chemistry. These two perturbative line lists were compiled into the conventional \EXOMOL{} format by \cite{Villanueva2018PlanetaryExoplanets} to form the previous most comprehensive line list for \NH{} across the UV to mid-IR regions.

There is no variational line list currently available for NH in the literature that has been constructed through numerical solution of the nuclear Schrödinger equation. This prevents reliable extrapolation to unobserved rovibrational states.

\subsection{\Abinitio{} data}\label{s:ab}
\Abinitio{} spectroscopic data have been calculated for many electronic states of NH to varying accuracy. The most~comprehensive studies were performed by \cite{OwonoOwono2007TheoreticalFunctions} and \cite{Song2016AccurateEffect}, which both consider many low-lying and high-lying Rydberg electronic states. Similar calculations were performed with fewer electronic states in other publications \citep{Li2021ExcellentInterfere,Brooke2014LineNH,Brooke2015Note:NH,18FernandoNH}. Despite this previous body of work, raw output data are generally not provided or important details, such as off-diagonal (transition) dipole moment curves are missing, which emphasises the importance of data availability as discussed in \cite{21McKemmishDiatomics}. Therefore, new \abinitio{} calculations were necessary for this study. 

In this work, \abinitio{} calculations were performed in Molpro 2020.1 \citep{Werner2020ThePackage} using the icMRCI-SD/aug-cc-pwCV5Z model chemistry with reference orbitals obtained from SA-CASSCF calculations. Using the ($a_1$, $b_1$, $b_2$, $a_2$) irreducible representation of the $C_{2v}$ point group, (3, 1, 1, 0) active orbitals from (N: $2s$, $2p$; H: $1s$) and (1, 0, 0, 0) closed orbitals from (N: $1s$) were selected. These orbitals were optimised for the \NHX{}, \NHA{}, \NHaa{}, \NHbb{} and \NHcc{} electronic states with equal weighting. These \abinitio{} calculations utilise the same model chemistry as  \cite{Brooke2014LineNH,Brooke2015Note:NH} and \cite{18FernandoNH}, but were expanded to include the singlet \NHaa{}, \NHbb{} and \NHcc{} electronic states.

\section[\NH{} Spectroscopic Model]{\NHbold{} Spectroscopic Model}\label{s:NHSM}
There is currently no available spectroscopic model (\textit{i.e.} a set of self-consistent potential energy, coupling and dipole moment curves used in the variational model of a molecule, see \cite{21McKemmishDiatomics} for details) fit to experimental data for \NH{} in the literature. The absence of this model hinders physically reasonable extrapolation to new unobserved transitions. 

Here, the \NH{} spectroscopic model is assembled for the five \NHX{}, \NHaa{}, \NHbb{}, \NHA{} and \NHcc{} electronic states in a \DUO{} \citep{16YurchenkoDuo} \MODEL{} input file. 

\subsection{Energy spectroscopic model}\label{s:NHesm}
The \NH{} energy spectroscopic model comprises the potential energy curve of each electronic state, as well as~the coupling and correction curves between these states. Firstly, the potential energy curves for the five electronic states considered are shown in Figure~\ref{f:PECNH}. The \NHbb{} and \NHcc{} curves were taken from the \abinitio{} data in Section~\ref{s:ab}. The \NHX{}, \NHaa{} and \NHA{} states were modelled using an extended Morse oscillator potential given by
\footnotesize
\begin{equation}\label{e:NHemo}
V(R) = T_e + D_e\left[1 - {\rm exp} \left(\sum^{N}_{i=0} b_i \left( \frac{R^4-R^4_e}{R^4+R^4_e} \right)^i \left(R_e-R\right)\right) \right]^2
\end{equation} 
\normalsize
where $T_e$ is the term energy, $D_e$ is the dissociation energy, $R_e$ is the equilibrium bond length and~$\{b_i$\} are vibrational~fitting parameters. The number of these fitting parameters differs on the left and right of the equilibrium bond length as specified by the parameters $N_L$ and $N_R$ in the \MODEL{} input file, respectively. The dissociation asymptote $A_e=T_e+D_e$ was set to 29,030~\cm{} \citep{Ram2010RevisedNH} for the \NHX{} electronic state and 48,256~\cm{} \citep{Brazier1986FourierNH} for the \NHA{} and \NHaa{} electronic states in accordance with \cite{18FernandoNH}.

\begin{figure}
\centering
\includegraphics[width=\linewidth]{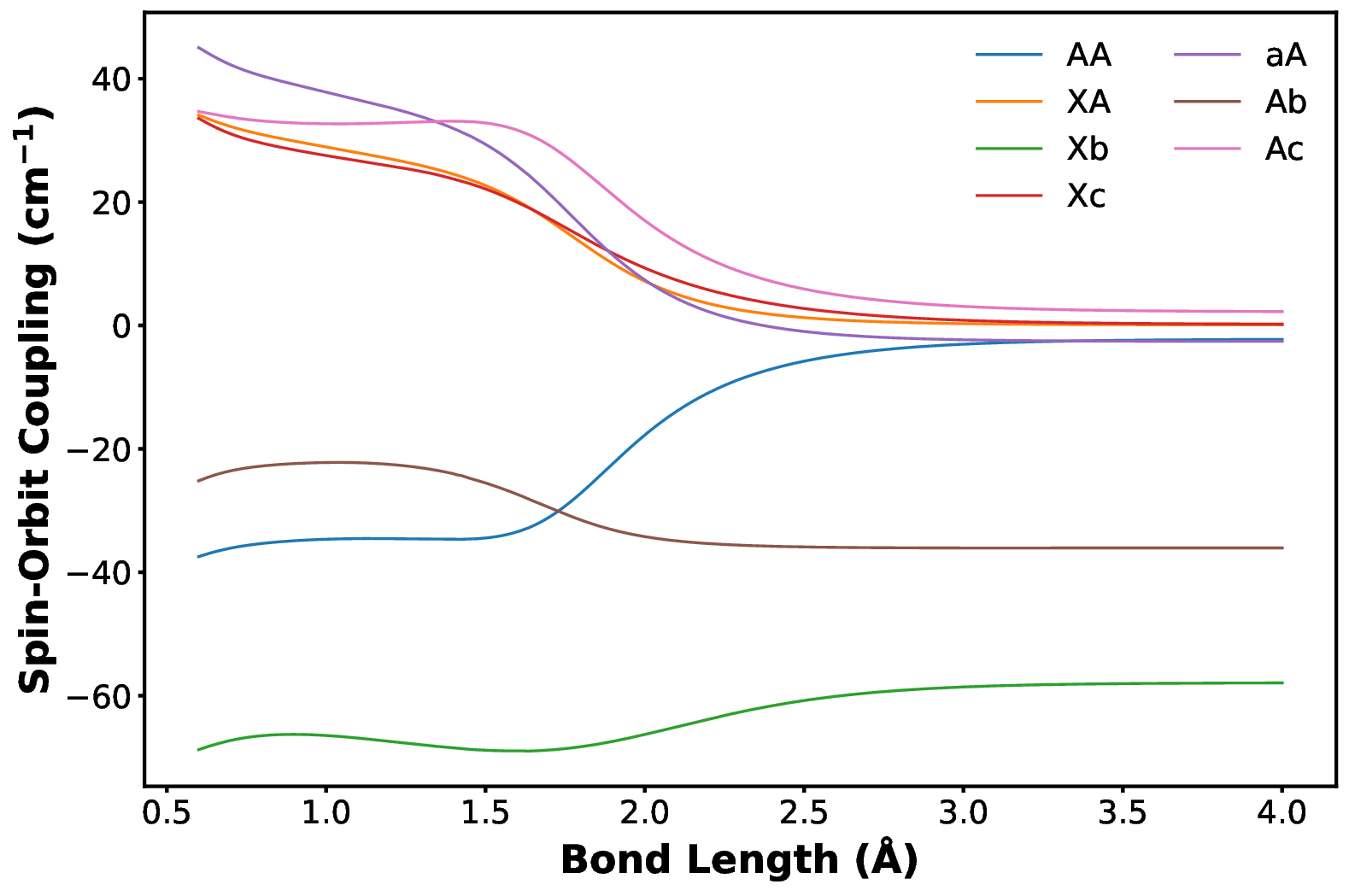}
\captionsetup{font=footnotesize}
\caption{The \NH{} \abinitio{} spin-orbit coupling curves calculated in this work.}
\label{f:so}
\end{figure}

The \NHX{}, \NHaa{} and \NHA{} potential energy curves were fit to \noMafitting{} unique \MARVEL{} experimental energies \citep{19DarbyNH}, where the inverse square of each \MARVEL{} uncertainty was applied as the weighting factor in the fitting procedure. The degenerate \MARVEL{} rovibronic states with unresolved $\Lambda$-doubling were duplicated with opposite parity for completeness. The \abinitio{} potential energy curves from Section~\ref{s:ab} were used to further confine the fit. The fitted parameters for the \NHX{}, \NHaa{} and \NHA{} potential energy curves are provided in Table~\ref{t:pecNH}.

\begin{table}
\centering
\caption{The fitted variational parameters of the potential energy and diagonal coupling curves for the \NHX{}, \NHaa{} and \NHA{} electronic states. Apart from the dimensionless parameters $N_L$~and $N_R$, all parameters are rounded to five significant figures in units of Å for $R_e$, Å$^{-i}$ for the $b_i$ vibrational parameters and \cm{} otherwise. The exact parameters employed can be found in the \MODEL{} file.}
\label{t:pecNH}
\begin{tabular}{cccc}
\hline
Parameter & \NHX{} & \NHaa{} & \NHA{} \\
\hline
$T_e$ & +0.0000E+00 & +1.2697E+04 & +2.9828E+04 \\
$R_e$ & +1.0374E+00 & +1.0340E+00 & +1.0385E+00 \\
$A_e$ & +2.9030E+04 & +4.8256E+04 & +4.8256E+04 \\
$N_L$ & 2 & 2 & 2 \\
$N_R$ & 6 & 4 & 4 \\
$b_0$ & +2.2744E+00 & +2.0838E+00 & +2.8059E+00 \\
$b_1$ & +5.1944E--02 & --3.7475E--02 & +3.7475E--01 \\
$b_2$ & +1.2097E--01 & +8.9429E--02 & +2.7966E--01 \\
$b_3$ & --1.1310E--01 & --1.5469E--01 & --4.6666E--02 \\
$b_4$ & +1.0407E+00 & +3.1844E--01 & +1.3859E+00 \\
$b_5$ & --1.8662E+00 & -- & -- \\
$b_6$ & +1.5602E+00 & -- & -- \\
\vspace{-0.5em} \\
$\lambda_\text{SS}$, $B_0$ & +5.3994E--01 & -- & --1.7225E--01 \\
$\lambda_\text{SS}$, $B_1$ & +6.3032E--01 & -- & --3.3004E--01 \\
$BO_\text{rot}$, $B_0$ & +2.3019E--03 & --4.9957E--04 & +2.2394E--03 \\
$\gamma_\text{SR}$, $B_0$ & +2.4946E--03 & -- & --7.7438E--04 \\
$\gamma_\text{SR}$, $B_1$ & +1.0706E--02 & -- & +2.6870E--02 \\
$\lambda_\text{p2q}$, $B_0$ & -- & -- & +1.3716E+00 \\
$\lambda_\text{opq}$, $B_0$ & -- & -- & --4.9803E--03 \\
$\lambda_\text{q}$, $B_0$ & -- & -- & --4.0109E--05 \\
\hline
\end{tabular}
\vspace{-3pt}
\end{table}

\begin{table}
\centering
\caption{A statistical summary of the \NH{} absolute energy residuals grouped by electronic state for each vibrational state $v$ in the \MARVEL{} analysis. The number of rovibronic states ($n$), as well as the root mean square error (RMSE), mean absolute error (MAE) and maximum residual in units of~\cm{}, are presented for each vibronic level.}
\label{t:resNH1}
\begin{tabular}{cccccc}
\hline
State & $v$ & $n$ & RMSE & MAE & Maximum\\
\hline
\multirow{7}{*}{\NHX{}}
& 0 & 105 & 1.641E--01 & 1.111E--01 & 4.533E--01 \\
& 1 & 103 & 2.232E--01 & 1.841E--01 & 3.787E--01 \\
& 2 & 97 & 1.773E--01 & 1.465E--01 & 3.052E--01 \\
& 3 & 86 & 2.784E--02 & 1.751E--02 & 1.107E--01 \\
& 4 & 64 & 8.715E--02 & 8.244E--02 & 1.833E--01 \\
& 5 & 51 & 6.191E--02 & 5.551E--02 & 1.019E--01 \\
& 6 & 36 & 1.753E--01 & 1.332E--01 & 4.415E--01 \\
\vspace{-0.5em} \\
\multirow{4}{*}{\NHaa{}}
& 0 & 36 & 1.339E--02 & 6.319E--03 & 5.873E--02 \\
& 1 & 38 & 1.819E--02 & 1.056E--02 & 7.775E--02 \\
& 2 & 24 & 4.362E--02 & 3.916E--02 & 7.690E--02 \\
& 3 & 12 & 4.901E--02 & 4.203E--02 & 8.717E--02 \\
\vspace{-0.5em} \\
\multirow{3}{*}{\NHA{}}
& 0 & 181 & 6.496E--02 & 4.455E--02 & 3.012E--01 \\
& 1 & 139 & 4.513E--02 & 2.955E--02 & 1.642E--01 \\
& 2 & 82 & 3.394E--02 & 2.946E--02 & 7.599E--02 \\
\vspace{-0.5em} \\
\multirow{2}{*}{\NHcc{}}
& 0 & 22 & 9.001E--01 & 6.370E--01 & 2.220E+00 \\
& 1 & 2 & 1.205E+01 & 1.205E+01 & 1.206E+01 \\
\hline
\end{tabular}
\vspace{-3pt}
\end{table}

Additionally, all spin-orbit ($A_\text{SO}$), spin-spin ($\lambda_\text{SS}$), spin-rotation ($\gamma_\text{SR}$), $\Lambda$-doubling ($\lambda_\text{opq}$, $\lambda_\text{p2q}$ and $\lambda_\text{q}$), angular momentum ($L_+$) and Born-Oppenheimer breakdown (BO$_\text{rot}$) interactions relevant to the \NHX{}, \NHaa{} and \NHA{} electronic states were considered in this energy spectroscopic model. Specifically, the Born-Oppenheimer breakdown correction allows for a position-dependent variation of the rotational mass \citep{LEROY2002175,2023BradySOll}, which is treated here for \NH{} only. This has shown to be important for hydrides with large rotational constants, such as NH \citep{C8CP04498H}, HF \citep{LEROY1999189} and MgH \citep{2013HendersonMgH}, in order to account for electronic centrifugal distortion. The spin-orbit and angular momenta curves obtained from the \abinitio{} calculations in \Cref{s:abinitio_NH} were inserted into the model without fitting to any functional form. The \abinitio{} spin-orbit curves are shown in Figure~\ref{f:so}. The remaining curves are purely empirical in nature and were fit to the \MARVEL{} experimental data using the damped polynomial given by
\footnotesize
\begin{equation}\label{e:Fr_NH}
    F(R)=\sum^{N}_{i=0}B_{i} \left[\left(R - R_e\right) e^{-(R-R_e)^2}\right]^i \left(1-\frac{R^2-R_e^2}{R^2+R_e^2}\right)
\end{equation}
\normalsize
These empirical interactions improve the overall quality of the fit, but decay to zero for large bond lengths in order to ensure good extrapolation. The diagonal coupling parameters are provided in Table~\ref{t:pecNH}.

Ultimately, the fitted energy spectroscopic model constructed herein can accurately reproduce the \MARVEL{} experimental energies for all electronic states. This is shown by the absolute energy residuals summarised statistically in \Cref{t:resNH1}, which were calculated as the absolute difference between the \MARVEL{} experimental and variational energies.

In \Cref{t:newparametersNH}, the \NHX{} and \NHA{} equilibrium spectroscopic parameters from this \NH{} model are compared to selected literature studies. There is a general agreement across all sources.

\begin{table*}
\centering
\captionsetup{font=footnotesize}
\caption{The \NH{} equilibrium spectroscopic parameters from selected sources reported without error estimates in units of Å for $R_e$ and \cm{} otherwise.}\label{t:newparametersNH}
\begin{tabular}{ccccccc}
\hline
State & Parameter & \LLname{} & \Abinitio{} & \citetalias{Huber1979MolecularMolecules} & \cite{Brazier1986FourierNH} & \cite{Song2016AccurateEffect} \\
\hline 
\multirow{7}{*}{\NHX{}} 
& $A_e$ & 29,030 & 28,341 & 27,987 & 29,030 & 29,153.70 \\
& $\omega_e$ & 3,281.8 & 3,279.1 & 3,282.27 & 3,282.220 & 3,292.07 \\
& $\omega_ex_e$ & 78.099 & -- & 78.35 & 78.513 & 86.66 \\
& $R_e$ & 1.0374 & 1.0369 & 1.03621 & -- & 1.0375 \\
& $B_e$ & 16.668 & 16.678 & 16.6993 & 16.667704 & 16.74 \\
& $D_e$ ($10^{-4}$) & 17.175 & 17.257 & 17.097 & -- & -- \\
& $\alpha_e$ & 0.64354 & 0.66548 & 0.649 & 0.649670 & 0.632 \\
\vspace{-0.5em} \\
\multirow{8}{*}{\NHA{}} 
& $T_e$ & 29,828 & 30,262 & 29,807.4 & 29,790.5 & 29,794.77 \\
& $A_e$ & 48,256 & 47,715 & -- & 48,256 & 48,186.62 \\
& $\omega_e$ & 3,225.8 & 3,238.7 & 3,231.2 & 3,231.70 & 3,263.32 \\
& $\omega_ex_e$ & 96.24 & -- & 98.6 & 98.48 & 97.73 \\
& $R_e$ & 1.0385 & 1.0369 & 1.03698 & -- & 1.0368 \\
& $B_e$ & 16.668 & 16.678 & 16.675 & 16.681963 & 16.69 \\
& $D_e$ ($10^{-4}$) & 17.666 & 17.690 & 17.800 & -- & -- \\
& $\alpha_e$ & 0.70933 & 0.72193 & 0.7454 & 0.712880 & 0.711 \\
\hline
\end{tabular}
\end{table*}

\subsection{Intensity spectroscopic model}\label{s:NHism}
\begin{figure}
\centering
\includegraphics[width=\linewidth]{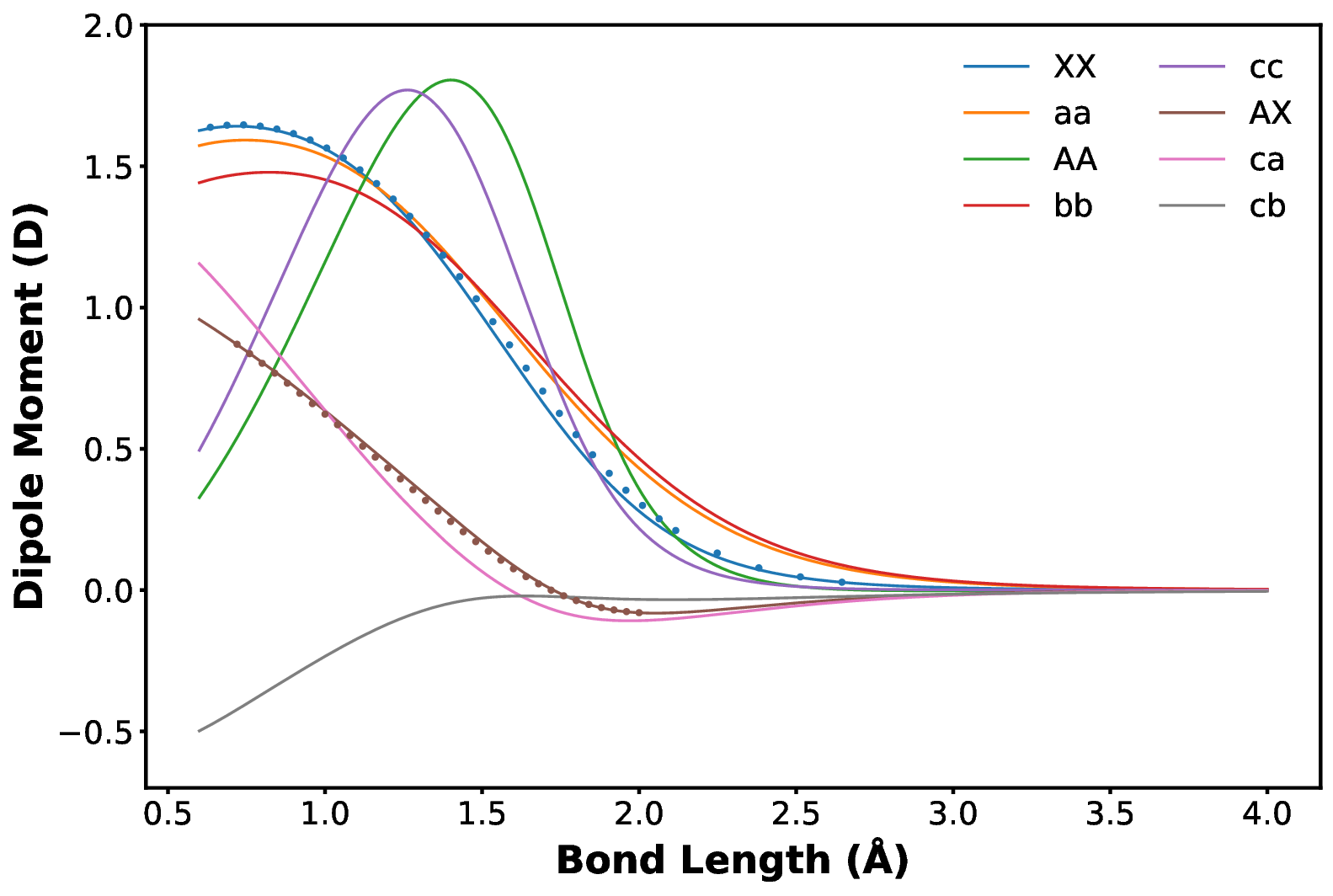}
\captionsetup{font=footnotesize}
\caption{The \NH{} \abinitio{} dipole moment curves calculated in this work. The coloured dots indicate values calculated by Brooke et. al. (2014, 2015)  and Fernando et. al. (2018), where a phase difference has been assumed for the \NHA{}~--~\NHX{} dipole moment curve.}
\label{f:dipole}
\end{figure}

The \NH{} intensity spectroscopic model contains only \abinitio{} dipole moment curves as is standard practice. These were inserted into the \MODEL{} file without fitting to a functional form, where all interpolation was performed by \DUO{} using natural splines. The intensity spectroscopic model includes all diagonal (permanent) and off-diagonal (transition) dipole moment curves shown in Figure~\ref{f:dipole}. This figure also includes values calculated by \cite{Brooke2014LineNH,Brooke2015Note:NH} and \cite{18FernandoNH} for the \NHX{}~--~\NHX{} and \NHA{}~--~\NHX{} curves, respectively. Their values are in close agreement with the new data; this is unsurprising given the electronic simplicity of NH and indicates that both dipole moment \abinitio{} data are of high quality.

\section{\NHl{} Line List Construction}\label{s:llc}
\subsection{Methodology overview}\label{s:mo}
The line list construction approach of \cite{Syme2021FullCN} and \cite{Perri2023} is followed herein for NH, where the \MARVEL{} experimental, \MOLLIST{} perturbative and \DUO{} variational energies are collated into a trihybrid line list. This methodology exploits the advantages of each methodology to produce \LLname{}, the most accurate and comprehensive line list for NH currently available.

Specifically, high accuracy is attributed to the high-resolution experimental \Marvel{} energies from \cite{19DarbyNH} complemented by a small number of interpolated perturbative (effective Hamiltonian) energies from \cite{Brooke2014LineNH,Brooke2015Note:NH,18FernandoNH}. High completeness arises from a new spectroscopic model (fitted to experimental data) constructed in the previous section that yields a variational line list. The experimental, perturbative and variational energies are combined to generate a novel trihybrid line list for \NH{}, named \LLname{}. Isotopologue energies for \isoa{}, \isob{} and \isoc{} are predicted from this main line list using standard techniques, where $^{14}$N, $^{15}$N, $^{1}$H and $^{2}$H have an isotopic abundance of 99.63, 0.37, 99.985 and 0.015 per cent, respectively \citep{Brown2003RotationalMolecules}.

In accordance with \EXOMOL{} line lists conventions, a trihybrid line list is prepared by updating the energies of the variational \STATES{} file. The trihybrid methodology prioritises the experimental energies due to their superior accuracy and reliable uncertainties. These replace the variational energies wherever possible with subsequent interpolation by the perturbative energies. The final trihybrid \STATES{} file includes the methodology and variational energy of each state, where the abbreviations Ma, EH and Ca represent the experimental (\MARVEL{}), perturbative (effective Hamiltonian) and variational (calculated) methodologies, respectively.

\begin{table*}
\caption{An extract from the \NH{} \STATES{} file. The total file is available at \url{www.exomol.com} and in the supplementary material.}
\label{tab:states}
\resizebox{\textwidth}{!}{
\begin{tabular}{cccccccccccccccccc}
\hline
$n$& $\tilde{E}$ & $g_{\rm tot}$ & $J$ & unc & $\tau$ & $g$ & $p_{+/-}$ & $p_\text{e/f}$ & State & $v$ & $\Lambda$ & $N$ & $F$ & Source & $\tilde{E}_\text{Ca}$\\
\hline
38 & 29829.555340 & 6 & 0 & 1.1500E--03 & 3.5430E--07 & 0.0000E+00 & - & f & A(3PI) & 0 & -1 & 1 & 3 & Ma & 29829.507210 \\
39 & 32862.788560 & 6 & 0 & 3.7136E--02 & 3.9583E--07 & 0.0000E+00 & - & f & A(3PI) & 1 & -1 & 1 & 3 & EH & 32862.755435 \\
40 & 35698.984598 & 6 & 0 & 7.5995E--02 & 4.4710E--07 & 0.0000E+00 & - & f & A(3PI) & 2 & -1 & 1 & 3 & Ca & 35698.984598 \\
41 & 38325.477243 & 6 & 0 & 6.0248E--01 & 5.1244E--07 & 0.0000E+00 & - & f & A(3PI) & 3 & -1 & 1 & 3 & Ca & 38325.477243 \\
\hline
\end{tabular}
}
\end{table*}

\begin{table}
\centering
\caption{The column descriptors of the NH \STATES{} files. Here,~$\hat{\mathbf{L}}$ is the electronic orbital angular momentum.}
\label{tab:statescd}
\resizebox{0.485\textwidth}{!}{
\begin{tabular}{ccc}
\hline
Column & Symbol & Descriptor \\
\hline
1 & $n$ & State Index \\
2 & $\tilde{E}$ & Energy (in \cm{}) \\
3 & $g_\text{tot}$ & Total Degeneracy \\
4 & $J$ & Total Angular Momentum \\
5 & unc & Uncertainty (in \cm{}) \\
6 & $\tau$ & Lifetime (in s) \\
7 & $g$ & Landé $g$-Factor \\
8 & $p_{+/-}$ & Total Parity \\
9 & $p_\text{e/f}$ & Rotationless Parity \\
10 & State & Electronic State\\
11 & $v$ & Vibrational State \\
12 & {$\Lambda$} & $\hat{\mathbf{L}}$ Projection Along Internuclear Axis \\
13 & $N$ & Rotational Angular Momentum \\
14 & $F$ & Fine Structure Counting Number \\
15 & Source & State Source (Ma, EH or Ca) \\
16 & $\tilde{E}_\text{Ca}$ & Variational (Calculated) Energy (in \cm{}) \\
\hline
\end{tabular}
}
\end{table}

\begin{table}
\caption{An extract from the \NH{} \TRANS{} file. The total file is available at \url{www.exomol.com} and in the supplementary material.}
\label{tab:trans}
\centering
\begin{tabular}{ccc}
\hline
$f$ & $i$ & $A_{fi}$ \\
\hline
83 & 1 & 4.5597E--02 \\
84 & 1 & 3.2919E+00 \\
85 & 1 & 8.7746E+00 \\
86 & 1 & 9.2748E--02 \\
\hline
\end{tabular}
\end{table}

\begin{table}
\centering
\caption{The column descriptors of the NH \TRANS{} files.}
\label{tab:transcd}
\begin{tabular}{ccc}
\hline
Column & Symbol & Descriptor \\
\hline
1 & $f$ & Final State Index \\
2 & $i$ & Initial State Index \\
3 & $A_{fi}$ & Einstein A Coefficient (in s$^{-1}$) \\
\hline
\end{tabular}
\end{table}

The \NH{} variational line list is output directly from \DUO{} using the energy and intensity spectroscopic model detailed in Section~\ref{s:NHSM}. The quantum numbers are given in terms of Hund's coupling case (a), which were converted to case (b). The \NH{} trihybrid line list was created by replacing the variational energies with \MARVEL{} experimental energies \citep{19DarbyNH} where available, followed by interpolation with \MOLLIST{} perturbative energies \citep{Brooke2014LineNH,Brooke2015Note:NH,18FernandoNH} to cover any missing experimental data. The variational energies of the \abinitio{} \NHcc{} electronic state were shifted here to match the \MARVEL{} analysis for the lowest energy level. These energy transformations were performed using an external Python script by grouping equivalent rovibronic states from each source by their electronic state, parity, $v$, $J$ and $N$ quantum numbers.

For all other isotopologues, a variational line list can be constructed trivially by changing the nuclear mass(es) in the main spectroscopic model. Following \cite{Polyansky2017ExoMolH217O}, the accuracy of this line list can be improved by applying an isotopologue extrapolation term that shifts each energy based on the difference between the variational and trihybrid line lists of the main \NH{} isotopologue as
\begin{equation}
\tilde{E}^\text{iso}_\text{IE} = \tilde{E}^\text{iso}_\text{Ca} + \left(\tilde{E}^\text{main}_\text{tri} - \tilde{E}^\text{main}_\text{Ca}\right)
\end{equation}
where $\tilde{E}^\text{main}_\text{Ca}$ is the calculated variational energy of the main isotopologue, $\tilde{E}^\text{main}_\text{tri}$ is its trihybrid energy, $\tilde{E}^\text{iso}_\text{Ca}$ is the calculated variational energy of a given isotopologue and $\tilde{E}^\text{iso}_\text{IE}$ is its isotopologue-extrapolated energy. In other words, it is assumed that the energy residuals between the hybridised and variational line lists are constant for all isotopologues.

NH is a light molecule that is best described using Hund's coupling case (b), and previous line lists have thus assigned quantum numbers under this regime. As Duo diagonalises the total rovibronic Hamiltonian using case (a) basis functions, however, the conversion to case (b) was a significant challenge faced in this work. For the \NHX{} electronic state, a simple empirical relationship was determined between the $N$ quantum number and the $\Omega=\Sigma$ quantum number assigned by \DUO{}. For the \NHA{} electronic state, an empirical relationship could not be determined that was consistent for all energy. The $N$ quantum number was thus assigned in an external Python script using based on energy orderings with comparison to existing line lists. For the singlet electronic states, there are no complications as $N=J$ in the absence of electronic spin, where $J$ is well-defined and always conserved. Despite these issues, it is important to note that \DUO{} solves the Schrödinger equation using a complete basis set and thus the energy of the quantum state is accurate despite the use of Hund's coupling case (a) in the calculation; indeed, the quantum numbers themselves (aside from $J$ and parity) are simply labels assigned based on dominant contribution to describe the wavefunction and have no influence on the final cross sections.

\subsection{Line list overview}
The \LLname{} trihybrid \STATES{} file contains \noenergies{} rovibronic states with \noMaenergies{} (\noMaenergiesunique{} unique) experimental states, \noEHenergies{} perturbative states and \noCaenergies{} variational states. These were calculated up to 47,500~\cm{}, which is just below the dissociation asymptote $A_e$ of the \NHaa{}, \NHbb{}, \NHA{} and \NHcc{} electronic states. A vibrational basis set up to $v$~=~60 was used for each electronic state with rotational states calculated to $J$~=~60. An extract of the final \STATES{} files is shown \Cref{tab:states} with associated column descriptors in \Cref{tab:statescd}.

The \LLname{} trihybrid \TRANS{} file contains \notransitions{}~transitions. All transitions were calculated up to 47,500~\cm{}. An extract of the final \TRANS{} files is shown \Cref{tab:trans} with associated column descriptors in \Cref{tab:transcd}. 

Additionally, isotopologue-extrapolated hybrid line lists were prepared for \isoa{}, \isob{} and \isoc{} using the technique in Section~\ref{s:mo}. In Table~\ref{tab:iso}, experimental \NHX{} rovibrational transition frequencies for all four isotopologues are compared statistically to the \LLname{} line list. These experimental transition frequencies were collated by \cite{C8CP04498H} from twelve sources and divided into rotational and infrared transitions with all hyperfine structure averaged over. There is excellent agreement for all isotopologues, where the largest residuals are observed for the \isoa{} infrared transitions. It should be noted that these isotopologue data are sparse with a maximum upper $v=6$, $J=27$ for the \isoa{} infrared transitions and $v=0$, $J=8$ for the \isob{} infrared transitions.

\subsection{Uncertainties}
The uncertainties assigned to the experimental, perturbative and variational energies were all evaluated differently. For the main \NH{} isotopologue, the experimental uncertainties were obtained directly from the \MARVEL{} analysis. The perturbative uncertainties were averaged from the experimental uncertainties with the same vibrational and rotational quantum numbers, where the experimental and perturbative energies are known to have similar accuracy when interpolating. The variational uncertainties were approximated using the energy residuals in \Cref{t:resNH1}. For energies within a vibrational state in the \MARVEL{} analysis, the uncertainty is given as the largest energy residual with the same vibrational quantum number. For energies within a vibrational state unknown by the \MARVEL{} analysis, the uncertainty is given as the twice the largest energy residual in the same electronic state. The \NHbb{} uncertainties were set to 100~\cm{}.

For the other three isotopologues, all \NHX{}, \NHaa{}, \NHA{} and \NHcc{} uncertainties were increased by an order of magnitude to account for errors arising from the isotopologue-extrapolation technique. The \NHbb{} uncertainties were kept at 100~\cm{}.

\begin{table}
\caption{A statistical summary of the absolute frequency residuals between \LLname{} and Melosso et. al. (2019) grouped by~isotopologue and transition type. The root mean square error (RMSE), mean absolute error (MAE) and maximum residual are presented in units of~\cm{}.}
\label{tab:iso}
\centering
\resizebox{0.485\textwidth}{!}{
\begin{tabular}{ccccc}
\hline
Isotopologue & Type & RMSE & MAE & Maximum \\
\hline
14N1H & Rotational & 2.910E--04 & 5.000E--04 & 1.275E--03 \\
14N1H & Infrared & 2.336E--03 & 3.618E--03 & 1.856E--02 \\
14N2H & Rotational & 4.498E--03 & 5.197E--03 & 9.616E--03 \\
14N2H & Infrared & 5.357E--01 & 5.550E--01 & 8.312E--01 \\
15N1H & Rotational & 6.986E--04 & 7.497E--04 & 1.175E--03 \\
15N1H & Infrared & 5.385E--04 & 8.674E--04 & 2.191E--03 \\
15N2H & Rotational & 4.081E--03 & 4.592E--03 & 8.361E--03 \\
\hline
\end{tabular}
}
\end{table}

\section{\NHl{} Line List Analysis}\label{s:lla}

\EXOCROSS{} \citep{Yurchenko2018EXOCROSS:Lists} was used to generate partition functions and cross sections for all four NH isotopologues studied here.

\begin{table*}
\centering
\caption{The partition functions for NH as a function of temperature $T$ (in K) from different sources. The main \NH{} isotopologue compares the data produced in this work to values calculated by \Mollist{} and Barklem \& Collet (2016).}
\label{t:pf}
\begin{tabular}{ccccccccc}
\hline
$T$ (K) & \NH & \NH & \NH & \isoa{} & \isob{} & \isoc{} \\
& \LLname{} & MoLLIST & \citetalias{Barklem2016} & \LLname{}& \LLname{} & \LLname{} \\
\hline
0 & 6.00000E+00 & 6.00000E+00 & 6.00000E+00 & 9.00000E+00 & 4.00000E+00 & 6.00000E+00 \\
1 & 1.80000E+01 & 1.80000E+01 & 1.80000E+01 & 2.70000E+01 & 1.20000E+01 & 1.80000E+01 \\
10 & 1.84915E+01 & 1.84915E+01 & 1.84904E+01 & 3.35570E+01 & 1.23345E+01 & 2.24658E+01 \\
100 & 8.29087E+01 & 8.29090E+01 & 8.29164E+01 & 2.23111E+02 & 5.54978E+01 & 1.49931E+02 \\
300 & 2.36350E+02 & 2.36350E+02 & 2.36370E+02 & 6.51909E+02 & 1.58246E+02 & 4.38188E+02 \\
500 & 3.90556E+02 & 3.90557E+02 & 3.90595E+02 & 1.08377E+03 & 2.61507E+02 & 7.28531E+02 \\
800 & 6.25183E+02 & 6.25184E+02 & -- & 1.75922E+03 & 4.18635E+02 & 1.18289E+03 \\
1,000 & 7.87584E+02 & 7.87585E+02 & 7.87722E+02 & 2.24860E+03 & 5.27422E+02 & 1.51233E+03 \\
1,500 & 1.23471E+03 & 1.23471E+03 & 1.23510E+03 & 3.67378E+03 & 8.27054E+02 & 2.47240E+03 \\
2,000 & 1.76222E+03 & 1.76209E+03 & 1.76315E+03 & 5.43911E+03 & 1.18069E+03 & 3.66229E+03 \\
3,000 & 3.11676E+03 & 3.11089E+03 & 3.12054E+03 & 1.01364E+04 & 2.08903E+03 & 6.82969E+03 \\
5,000 & 7.36897E+03 & 7.13095E+03 & 7.40652E+03 & 2.53046E+04 & 4.94084E+03 & 1.70616E+04 \\
\hline
\end{tabular}
\end{table*}

\subsection{Partition functions}

The partition functions $Q(T)$ were calculated using
\begin{equation}
Q(T) = \sum_n g_n^\text{ns}(2J_n+1)e^{-c_2\tilde{E}_n/T}
\end{equation}
where $g^\text{ns}$ is the nuclear spin statistical weight factor, $J$ is the total angular momentum excluding nuclear spin, $c_2 = hc/k_B$ (in cm K), $\tilde{E}$ is the energy term value (in cm$^{-1}$) and $T$ is temperature (in K). 

In \Cref{t:pf}, the four partition functions are evaluated at selected temperatures. For \NH{}, the partition function is compared to values calculated by \MOLLIST{} and \cite{Barklem2016}. There is strong agreement between the \LLname{} data and \cite{Barklem2016}, even at 5,000~K, thus strongly corroborating both data sets. There is also strong agreement with the \MOLLIST{} data, where the differences at 5,000~K arise due to the inclusion of higher vibrational bands from the variational calculations. Therefore, the \LLname{} trihybrid line list provides only marginal improvements to the partition functions and is primarily included for line list completeness and ease of access.

\subsection{Cross sections}\label{s:NHacs}
\begin{figure*}
\centering
\includegraphics[width=\linewidth]{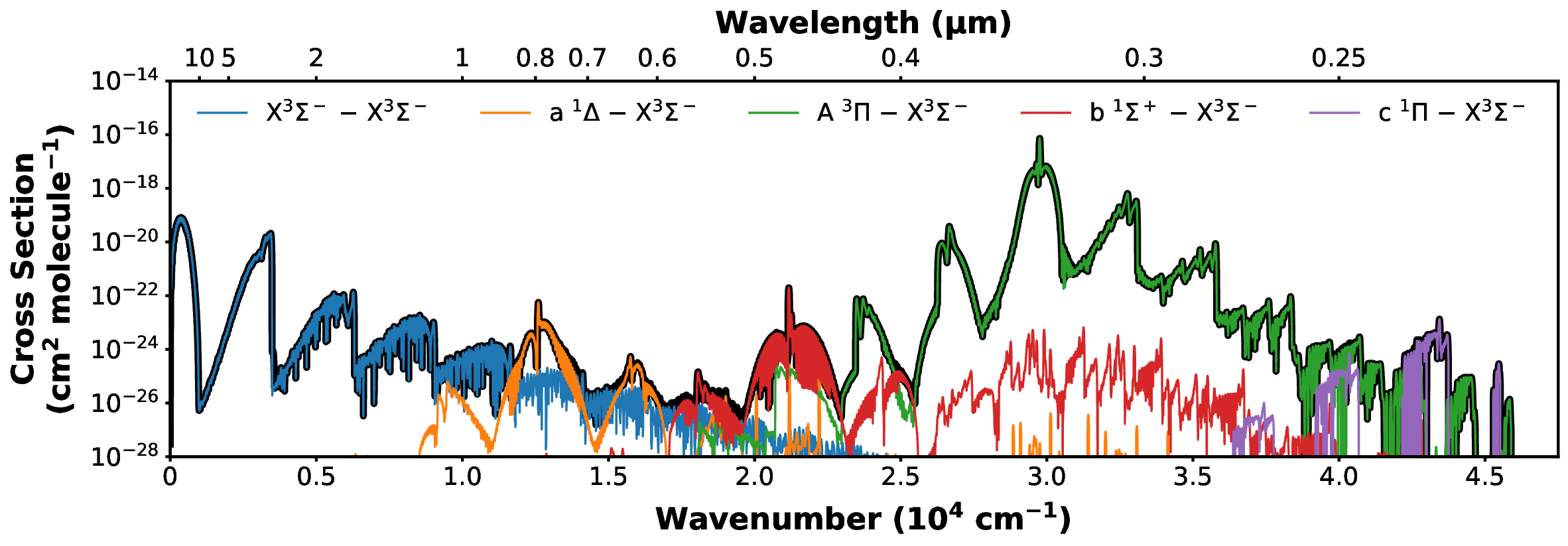}
\caption{The total \NH{} absorption cross section decomposed for 0~--~47,500~\cm{} at 2,000~K.}
\label{f:xsec_all_1_NH}
\vspace{12pt}
\end{figure*}

\begin{figure*}
\centering
\includegraphics[width=\linewidth]{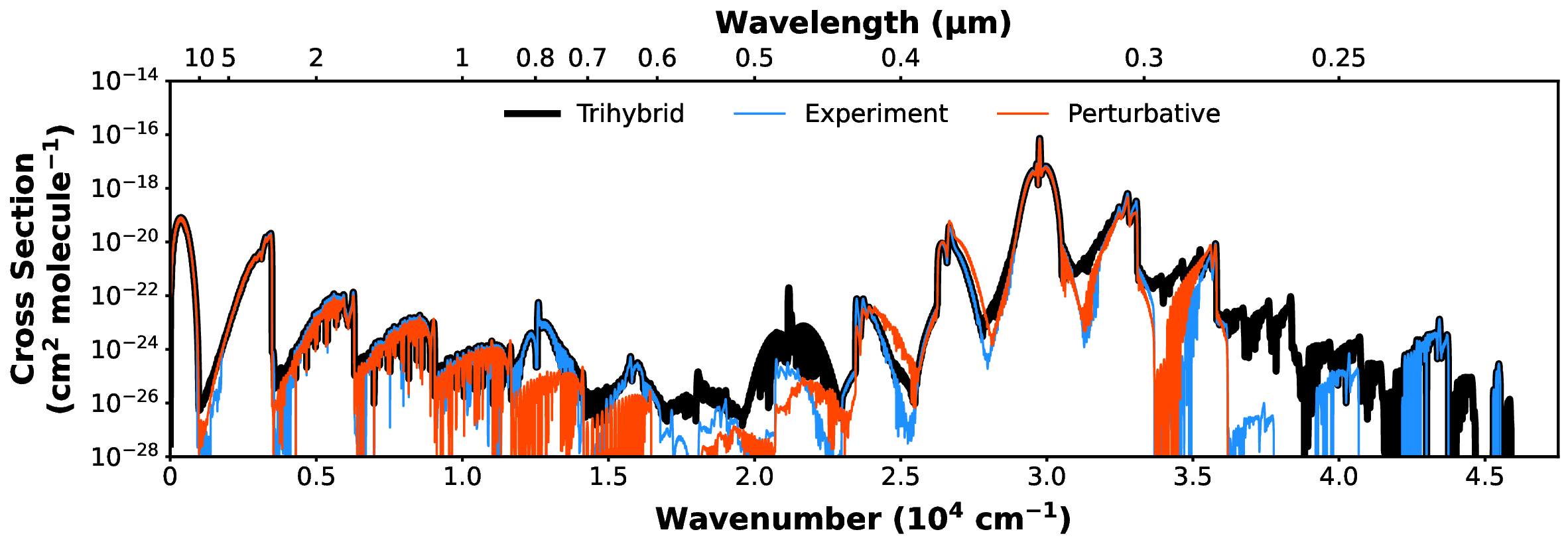}
\caption{The total \NH{} absorption cross section simulated at 2,000 K from different sources. The black, blue and orange cross~sections show transitions accounted for by the \LLname{} trihybrid line list, \MARVEL{} experimental line list by Darby-Lewis et. al. (2019)  (with variational intensities) and \MOLLIST{} perturbative line list by Brooke et. al. (2014,2015); Fernando et. al. (2018), respectively.}
\label{f:xsec_all_3_NH}
\vspace{12pt}
\end{figure*}

\begin{figure*}
\centering
\includegraphics[width=\linewidth]{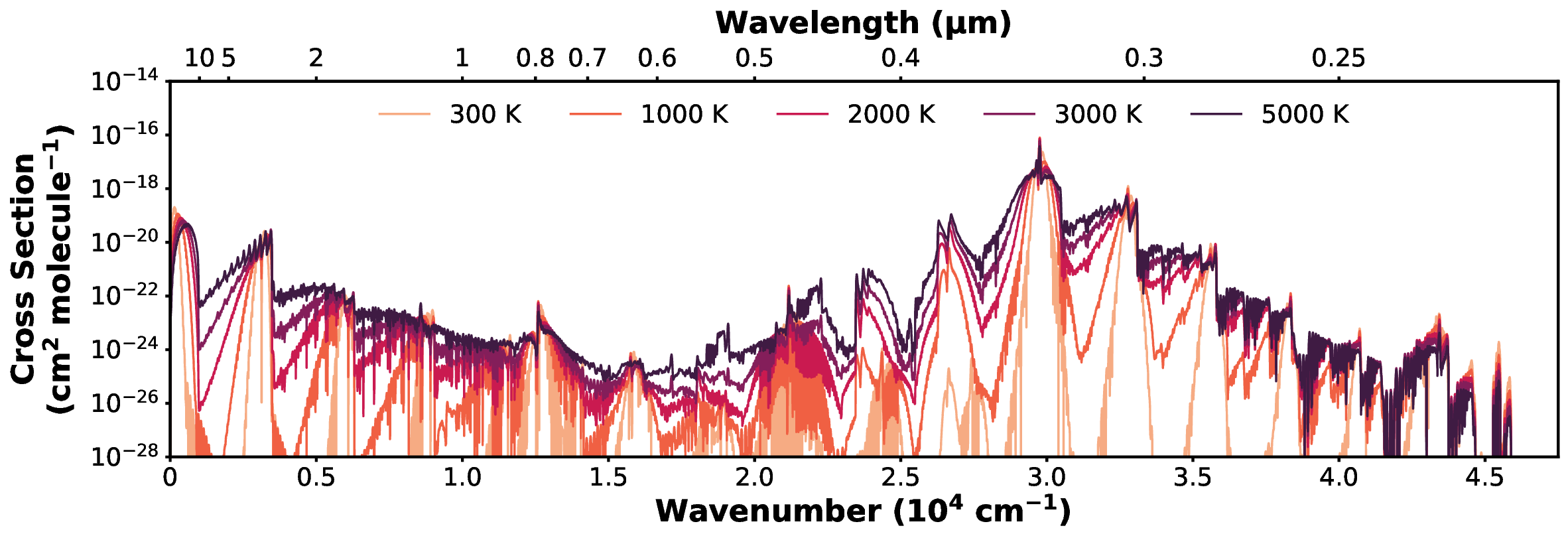}
\caption{The total \NH{} absorption cross section simulated at several temperatures.}
\label{f:xsec_all_4_NH}
\end{figure*}

Here, cross sections are presented for the main \NH{} isotopologue only. The absorption cross sections in Figures \ref{f:xsec_all_1_NH}, \ref{f:xsec_all_3_NH} and \ref{f:xsec_all_4_NH} were simulated using a Gaussian lineshape with a half-width at half-maximum of 20~\cm{} (artificially large to aid visibility; NH will not exhibit significant Doppler broadening in its common astronomical environments).

In Figure~\ref{f:xsec_all_1_NH}, the total \NH{} absorption cross section at 2,000 K is decomposed into its primary transition bands for 0~--~47,500~\cm{}. The spectrum possesses many intense transitions in all spectral regions. The ultraviolet region is the most prominent, where the \NHA{} -- \NHX{} transitions dominate. The spin-forbidden \NHaa{} -- \NHX{} and \NHbb{} -- \NHX{} systems are observable over the \NHX{} rovibrational system in the visible region.

In Figure~\ref{f:xsec_all_3_NH}, the total \NH{} absorption cross section at 2,000~K is generated using three line lists. The black, blue and orange cross sections indicate transitions provided by the trihybrid, \MARVEL{} experimental (with variational intensities) and \MOLLIST{} perturbative line lists, respectively. As expected, the trihybrid line list produces the most comprehensive cross section due to its inclusion of the variational transition intensities calculated in this work. The \MARVEL{} analysis offers good coverage of the \NHA{}~--~\NHX{} and \NHaa{}~--~\NHX{} rovibronic transitions, as well as the \NHX{} rovibrational system. The \MOLLIST{} analysis agrees well with the trihybrid cross section for the main vibrational bands of the \NHA{}~--~\NHX{} manifold. The \NHaa{}~--~\NHX{} and \NHbb{}~--~\NHX{} rovibronic transitions are not included in the \MOLLIST{} line lists.

In Figure~\ref{f:xsec_all_4_NH}, the total \NH{} absorption cross section is simulated at several temperatures. The weaker bands in the cross section increase in intensity and all bands become broader with some shape changes with increasing temperature. 

\subsection{High-resolution considerations}
In order to assess the suitability of a line list for use in the HRCC detection of a molecule, it is critical to know the accuracy with which line positions are known. The accuracy of a transition frequency depends on the source of the energies of its upper and lower state. If both the upper and lower state energies are known from \MARVEL{} data (with typically low uncertainties), then its frequency can be assumed to be known to high accuracy. Conversely, if one or both of the energies are calculated from the Duo variational spectroscopic model, then the frequency will have much higher uncertainty; it is assumed here that it is insufficiently accurate for HRCC studies. The effective Hamiltonian perturbative data can be neglected from this discussion due to the low number of energies in the \LLname{} line list.

Again, in Figure~\ref{f:xsec_all_3_NH}, the cross section of the full \LLname{} trihybrid line list (assumed to be near complete based on partition function considerations and the inclusion of all low-lying electronic states) is compared with the cross section predicted solely by \MARVEL{} data (assumed to be highly accurate). In spectral regions where the two curves overlap, it should be understood that the transition frequency of the strong line is predicted to experimental (\Marvel{}) accuracy. In this case, Figure~\ref{f:xsec_all_3_NH} shows that the \LLname{} trihybrid line list has sufficient accuracy to be suitable for HRCC primarily between 22,500~--~35,000~\cm{} (280~--~444~nm) and below 15,000~\cm{} (above 667~nm) at 2,000~K. Despite this, many spectral lines are relatively weak or in spectral regions that are difficult to observe, such as the strong UV lines near 30,000~\cm{} (333~nm). The most easily observed transitions are probably the N~--~H fundamental stretch vibrational band near 3,300 \cm{} (3.03 \um{}). These conclusions are based on the 2,000~K cross section, but are readily generated at different temperatures using \EXOCROSS{} and can be examined at higher resolutions to ensure suitability for a desired application.

\begin{figure}
\centering
\includegraphics[width=\linewidth]{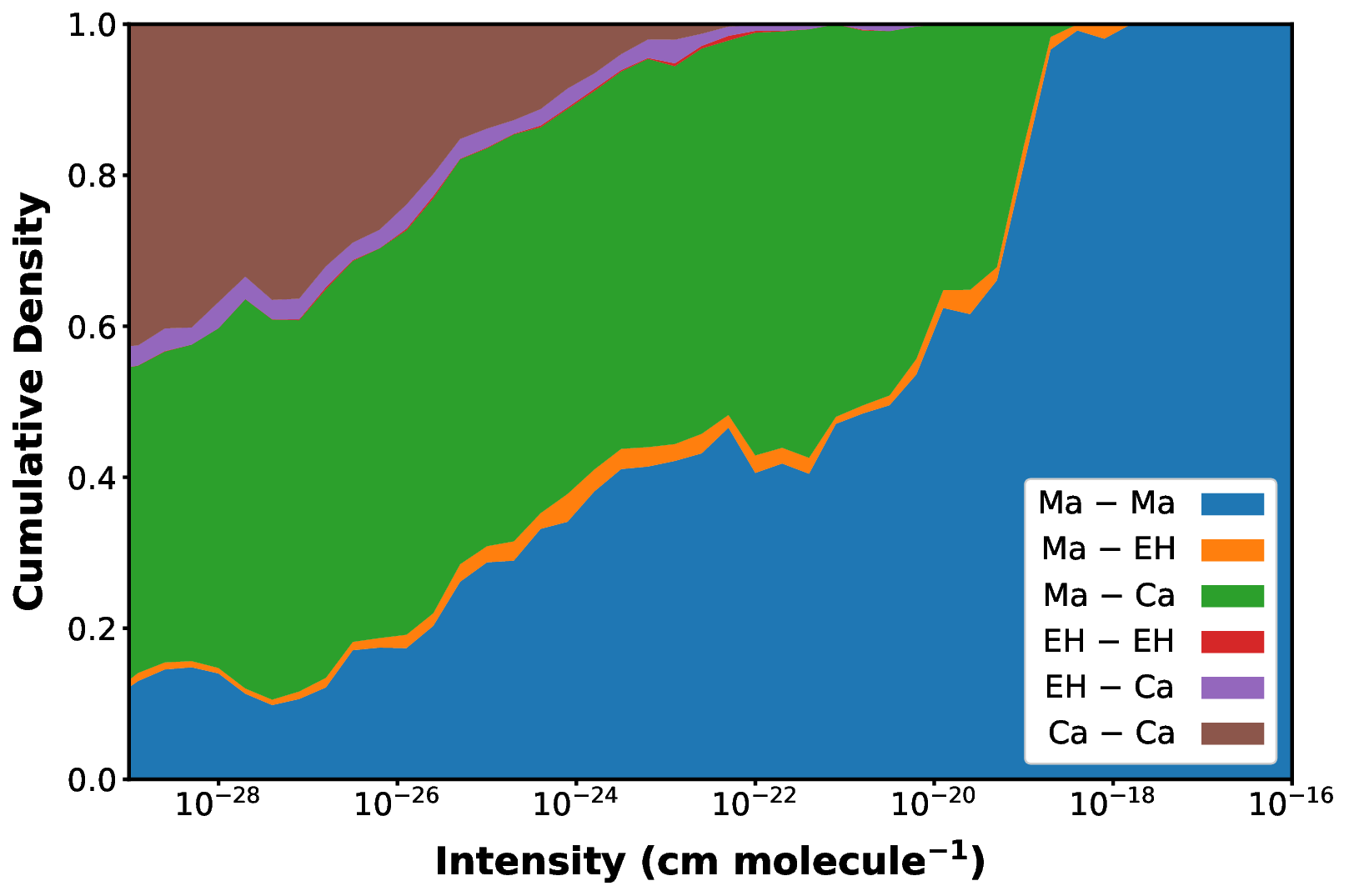}
\caption{The cumulative density of energy sources as a function of transition intensity at 2,000~K. The abbreviations Ma, EH and Ca represent the experimental (MARVEL), perturbative (effective Hamiltonian) and variational (calculated) methodologies, respectively.}
\label{f:intdis}
\end{figure}

A useful complementary perspective on high-resolution completeness is obtained from Figure~\ref{f:intdis}. This plot is read~by looking at vertical slices to visualise the source of upper~and lower state energies for transitions above that intensity. For example, the vertical slice at $10^{-20}$ cm molecule$^{-1}$ shows that about 60 per cent of all transitions with intensity of $10^{-20}$ cm molecule$^{-1}$ have both upper and lower energies from the high-resolution \MARVEL{} data; however, this still leaves about 40 per cent of lines at 2,000~K whose positions are not sufficiently accurate for HRCC. 

Despite the high accuracy of the data for high-resolution cross-correlation, it must be noted that for NH specifically, most of the strong bands are in challenging spectral regions for ground-based observations and thus the high-resolution cross-correlation techniques are likely to face considerable challenges. Space-based observations are thus highly preferable for detecting NH.

\subsection{Discussion and future direction}
For the anticipated applications of NH spectral data in all regions except the far-IR, the \LLname{} line list should provide information of sufficient quality. It should not be used for high-accuracy studies in the far-IR region (\textit{i.e.} the rotational spectroscopy) because, in this spectral region, there is significant hyperfine structure that are not incorporated in this spectroscopic model and thus not present in this line list. Specifically, hyperfine interactions are due to the nuclear spins of $^{14}$N ($I=1$) and $^1$H ($I=1/2$) for \NH{}. This hyperfine structure is resolved in most rotational studies, but not for any of the higher wavenumber studies \citep{19DarbyNH}. With this limitation in experimental data, at this stage, the best line list data source for modelling the rotational (far-IR) spectroscopy of \NH{} remains distinct from the best line list data source of the infrared-UV absorption (\LLname{}, developed here). For the far IR observations, it is recommended that astronomers use line list data from Splatalogue \citep{Splatalogue}, which combines both JPL \citep{JPL} and CDMS \citep{CDMS} data for NH in one convenient online interface. It should be noted that an extended version of \Duo{} has just been released \citep{HyperfineDuo} that does allow hyperfine-resolved line lists to be generated (although for only one nucleus with nuclear spin); this could be a future avenue for the improvement of NH data and could potentially allow a unified single source of spectroscopic data across the electromagnetic spectrum.

Any improvements in the accuracy of the existing non-hyperfine-resolved line list should focus on more precise quantification of the parameters relevant to forbidden transitions in the visible spectral region. This can be accomplished by improvements in understanding the energies of the \NHbb{} state (\textit{e.g.} from high-resolution spectroscopy experiments) and the \NHaa{}~--~\NHA{} and \NHbb{}~--~\NHA{} spin-orbit couplings (theoretical models may actually be a better approach here to avoid complications from other factors). The current \LLname{} trihybrid line list demonstrates that these spin-orbit coupling terms are the physical cause of the strongest transitions in the spectral region from 450~--~800~nm, the spin-forbidden \NHaa{}~--~\NHX{} and \NHbb~--~\NHX{} transitions. The current model should have semi-quantitative accuracy as the reliability of \abinitio{} calculations for NH are high. That said, experimental verification would be welcome.

\section{Conclusions}
The best current understanding of NH spectroscopy across the infrared to UV regions is presented herein through the novel \LLname{} trihybrid line list, which was produced from a new spectroscopic model containing five low-lying electronic states \NHX{}, \NHA{}, \NHaa{}, \NHbb{} and \NHcc{} for all major isotopologues (\NH{}, \isoa{}, \isob{} and \isoc{}). In order to ensure the highest accuracy of line positions currently possible, experimental (\Marvel{}; \cite{19DarbyNH}) and perturbative (\Mollist{}; \cite{Brooke2014LineNH,Brooke2015Note:NH,18FernandoNH}) energies have been incorporated into both the spectroscopic model fitting procedure and the final trihybrid line list. In order to ensure very high completeness, novel \abinitio{} data were utilised to create a full variational spectroscopic model. It should be cautioned that the Splatalogue data \citep{Splatalogue} for NH are preferable for far-IR studies where available, given that these data are hyperfine resolved. The resulting hyperfine line splittings are important for the accurate modelling of rotational spectral lines (observed in far-IR studies), although these have thus far not been resolved in the mid-IR to UV spectral regions.

The strong NH absorption lines are unfortunately in inconvenient spectral ranges for ground-based telescopes (UV peaking at 335~nm, mid-IR near water absorption and far-IR), and thus molecular detection via high-resolution cross-correlation techniques will likely be very challenging. Nevertheless, the molecular data for NH is sufficient to allow these investigations from 280~--~440~nm and above 667~nm. Space-based telescopes are thus the preferred way of detecting NH in astronomical environments and the new \LLname{} line list will support astronomers in identifying NH and quantifying its abundance using telescopes across a variety of spectral regions, including UV, visible, near and mid-IR.  

\section*{Acknowledgements}
This research was undertaken with the assistance of resources from the National Computational Infrastructure (NCI Australia), an NCRIS enabled capability supported~by the Australian Government.

The authors declare no conflicts of interest. 

The authors thank Charlie Drury, Sergey Yurchenko and Jonathan Tennyson for sharing their insights into the astronomical importance of NH. 

\section*{Data Availability}
The most up-to-date \LLname{} line list data for NH are available on the ExoMol website (\url{www.exomol.com}), specifically: 
\begin{itemize}[leftmargin=*]
\item The main \LLname{} \MODEL{} file (14N1H\_kNigHt.model).
\item The main \LLname{} partition function from 0 -- 10,000 K (14N1H\_kNigHt.pf).
\item The main \LLname{} \STATES{} file (14N1H\_kNigHt.states).
\item The main \LLname{} \TRANS{} file (14N1H\_kNigHt.trans).
\item All other isotopologue \MODEL{}, .pf, \STATES{} and \TRANS{} files (in their respective folders).
\end{itemize}

These data at the time of publication are also available as supporting information for this paper. The \MOLPRO{} quantum chemistry input files for the MRCI calculations are also included to assist data reproducibility.

\balance


\end{document}